\documentclass[aps,prb,twocolumn,superscriptaddress]{revtex4}

\usepackage[dvips]{graphicx}
\usepackage[dvips]{color}
\usepackage{amsmath}
\usepackage{amssymb}
\usepackage{multirow}

\usepackage{bm}

\def\t#1{\textrm{#1}}
\def\n{\nonumber \\ }
\def\tensor{\otimes}

\begin{document}

\title{Stability of surface states of weak $\mathbb{Z}_2$ topological insulators and superconductors}

\author{Takahiro Morimoto}
\affiliation{Condensed Matter Theory Laboratory, RIKEN, Wako, Saitama, 351-0198, Japan}
\author{Akira Furusaki}
\affiliation{Condensed Matter Theory Laboratory, RIKEN, Wako, Saitama, 351-0198, Japan}
\affiliation{RIKEN Center for Emergent Matter Science (CEMS), Wako, Saitama, 351-0198, Japan}
\date{\today}

\begin{abstract}
We study the stability against disorder of surface states of
weak $\mathbb{Z}_2$ topological insulators (superconductors)
which are stacks of strong $\mathbb{Z}_2$ topological insulators
(superconductors),
considering representative Dirac Hamiltonians
in the Altland-Zirnbauer symmetry classes
in various spatial dimensions.
We show that, in the absence of disorder,
surface Dirac fermions of weak $\mathbb{Z}_2$
topological insulators (superconductors)
can be gapped out by a Dirac mass term 
which couples surface Dirac cones and
leads to breaking of a translation symmetry (dimerization).
The dimerization mass is a unique Dirac mass term in
the surface Dirac Hamiltonian, and
the two dimerized gapped phases which differ in the sign of the Dirac
mass are distinguished by a $\mathbb{Z}_2$ index.
In other words the dimerized surfaces can be regarded as a strong
$\mathbb{Z}_2$ topological insulator (superconductor).
We argue that the surface states are not localized by disorder
when the ensemble average of the Dirac mass term vanishes.
\end{abstract}

\pacs{72.10.-d,73.20.-r,73.43.Cd}

\maketitle


\section{Introduction}

Three-dimensional topological insulators (TIs)\cite{hasan-kane10,qi-zhang-rmp11,fu-kane-mele07,fu-kane-inv07,moore-balents07,roy-3dTI09}
are classified into strong and weak TIs.
A three-dimensional strong TI is characterized by an
intrinsically three-dimensional topological index
(strong $\mathbb{Z}_2$ index $\nu_0$) and
has an odd number of gapless Dirac cones on every surface.
By contrast, a three-dimensional weak TI
is adiabatically connected to stacked layers of two-dimensional TIs,
characterized with three weak topological indices $(\nu_1,\nu_2,\nu_3)$
specifying the stacking direction,
and has an even number
(typically two) of Dirac cones on its side surfaces.
Since an even number of Dirac cones can be gapped out without breaking
time-reversal symmetry, it was initially considered
that gapless surface states of weak TIs are fragile.

However, recent theoretical studies have revealed unexpected strength of
weak TIs.
It was first argued by Ringel \textit{et al.}\cite{Ringel}\ that
surface Dirac fermions of weak TIs are not localized by
random potential which is weaker than a band gap and has zero mean.
This was confirmed by a numerical study of
effective Hamiltonian of two Dirac cones on the surface of a weak TI
perturbed by various
disorder potentials which preserve time-reversal symmetry.\cite{Mong}
The surface Dirac Hamiltonian used in this simulation
has $4\times4$ matrix representation and has a single Dirac mass
term which physically represents dimerization of stacked layers.%
~\cite{fu-kane-inv07,Ringel,Mong,liu12,FuKane2012}
Mong \textit{et al.}\ have also shown that a nonvanishing average of the
Dirac mass induces a transition from a metallic phase to an insulating
phase.\cite{Mong}
A more recent numerical study using a network model
has shown that the metal-insulator transition in the presence of
a finite dimerization belongs to the standard universality class of
two-dimensional symplectic class of Anderson localization.\cite{Obuse}

The remarkable stability against disorder of surface states of undimerized
weak TIs is ascribed to the uniqueness of the dimerization
mass term in the effective Hamiltonian for the surface states.%
~\cite{Mong,FuKane2012}
The point is that
the sign of the unique Dirac mass term distinguishes topologically
distinct phases,
since the effective Hamiltonian has the same form as the
Bernevig-Hughes-Zhang model\cite{Bernevig2006} of two-dimensional
quantum spin Hall insulators.
This then leads to the following semiclassical picture of
surface transport.
When random potential (including the Dirac mass) varies slowly in space,
the two-dimensional surface of a weak TI
is divided into positive-mass domains 
and negative-mass domains. 
Every domain boundary has helical gapless modes, which will percolate
over the surface when positive- and negative-mass domains
appear with the same probability, i.e., when the random potential
has zero mean.
As pointed out by Fu and Kane,\cite{FuKane2012}
this physics is similar to integer quantum Hall plateau transitions.
It is well known that the two-dimensional Dirac Hamiltonian
$H=k_x\sigma_x+k_y\sigma_y+m\sigma_z$
exhibits anomalous quantum Hall effect
$\sigma_{xy}=-\mathrm{sgn}(m)e^2/2h$, where $\sigma_i$ are Pauli matrices
and $k_x$ and $k_y$ are momenta.
The critical point between the quantum Hall plateaus $\sigma_{xy}=\pm e^2/2h$
is realized by tuning ensemble average of the random mass $m$
to zero.\cite{Nomura2008}
Similarly, a clean surface of an undimerized weak $\mathbb{Z}_2$ TI
is exactly at a quantum critical point separating two dimerized insulating
phases which are distinguished by a $\mathbb{Z}_2$ index.
A weak disorder then turns the quantum critical point
into a metallic phase (because of anti-localization),
as observed in numerical simulations.\cite{Mong}

The purpose of this paper is to show that the stability of surface
states of weak TIs against disorder discussed above can be generalized to
much broader class of weak TIs and topological superconductors (TSCs)
of any symmetry class which are
characterized by weak $\mathbb{Z}_2$ indices.
It is well known that, in every spatial dimension,
two out of the ten Altland-Zirnbauer symmetry
classes\cite{AZ-classes} of noninteracting fermion systems
can have topologically nontrivial gapped phases of noninteracting fermions
characterized by a strong $\mathbb{Z}_2$ index.%
~\cite{schnyder-ryu08,kitaev09,ryu-njp10}
For example, symmetry class AII, a set of free-fermion Hamiltonians
which are invariant under time-reversal operation $T$ with $T^2=-1$,
has two- and three-dimensional TIs
with a strong $\mathbb{Z}_2$ index.\cite{hasan-kane10,qi-zhang-rmp11}
In this paper we shall study $(d+1)$-dimensional weak $\mathbb{Z}_2$ TIs
(TSCs)
which are stacks of $d$-dimensional strong $\mathbb{Z}_2$ TIs (TSCs).
As a representative model for $d$-dimensional strong $\mathbb{Z}_2$
TIs (TSCs), we take a $d$-dimensional Dirac Hamiltonian
satisfying symmetry conditions of a given symmetry class.
We will show that a side surface of stacks of $d$-dimensional
strong $\mathbb{Z}_2$ TIs (TSCs) has two Dirac cones
which are again described by a $d$-dimensional Dirac Hamiltonian.
It admits a unique Dirac mass term which couples the two Dirac cones
and opens a gap.
To this end, we will employ Clifford algebras to treat a Dirac Hamiltonian
and symmetry constraints on equal footing and make use of topological
properties of classifying spaces which in our case are sets of all possible
Dirac mass terms.
The sign of the unique Dirac mass corresponds to a $\mathbb{Z}_2$
topological index of insulating phases.
We will then argue that surface Dirac fermions cannot be localized
by weak random potential if disorder average of the Dirac mass
term vanishes.

The organization of this paper is as follows.
In Sec.\ II, we briefly review
the classification of TIs and TSCs using relevant Clifford algebras.
In Sec.\ III, we describe a model of weak $(d+1)$-dimensional TIs/TSCs
which are stacks of strong $\mathbb{Z}_2$ TIs/TSCs for real symmetry classes,
and derive an effective Dirac Hamiltonian for surface states
with a unique Dirac mass term.
In Sec.\ IV, we discuss a couple of examples of weak $\mathbb{Z}_2$
TIs and TSCs.
We explicitly show that the mass term gapping out the surface states
is of dimerization type and unique.
In Sec.\ V, we conclude by summarizing the main results.

\section{Minimal Dirac model and Clifford algebras}

In order to introduce theoretical formalism which we employ
in the following sections, we give a brief review of
Clifford algebras and their application to classification
of TIs and TSCs.

We consider a Dirac Hamiltonian
\begin{align}
H=\gamma_0 m+\sum_{i=1}^d \gamma_i k_i,
\label{eq:Dirac-H}
\end{align}
where $\gamma_i$'s ($i=0,1,\ldots,d$) are gamma matrices anticommuting
with each other,
\begin{equation}
\{\gamma_i,\gamma_j\}=2\delta_{i,j},
\end{equation}
$k_i$ is a momentum in the $i$th direction,
and $m$ is the Dirac mass that corresponds to a band gap.\cite{ryu-njp10}
We have set the velocity of Dirac fermions to unity.
The Dirac Hamiltonian [Eq.~(\ref{eq:Dirac-H})] is a minimal (irreducible) model,
as we assume that no unitary operator (generator of continuous
symmetry such as rotation) commutes with $H$.
Such a Hamiltonian is classified as a member of one of the ten
Altland-Zirnbauer (AZ) symmetry classes,\cite{AZ-classes}
according to the presence or absence of the three generic symmetries:
time-reversal symmetry (TRS), particle-hole symmetry (PHS),
and chiral symmetry.
The Hamiltonian $H$ has a TRS when it commutes with an antiunitary
operator $T$ for time-reversal transformation as
\begin{subequations}
\label{eq:TRS}
\begin{equation}
[T,H]=0,
\label{eq:TRSa}
\end{equation}
which implies
\begin{align}
\{T,\gamma_i\} &=0 \quad(\t{for }i\neq 0), &
[T,\gamma_0] &=0,
\label{eq:TRSb}
\end{align}
\end{subequations}
for the Dirac Hamiltonian $H$,
because complex conjugation operator $\mathcal{K}$
involved in $T$ changes $k_i$ to $-k_i$.
Similarly, the Hamiltonian $H$ has a PHS when it anticommutes with an
antiunitary operator $C$,
\begin{subequations}
\label{eq:PHS}
\begin{equation}
\{C,H\}=0,
\label{eq:PHSa}
\end{equation}
or
\begin{align}
[C,\gamma_i] &=0 \quad(\t{for }i\neq 0), &
\{C,\gamma_0\} &=0,
\label{eq:PHSb}
\end{align}
\end{subequations}%
where $C$ serves as an operator for particle-hole transformation
of Bogoliubov-de Gennes (BdG) Hamiltonian of a superconducting system.
The Hamiltonian $H$ has a chiral symmetry, if there exists
a unitary operator $\Gamma$ that anticommutes with $H$,
\begin{subequations}
\label{eq:chiral}
\begin{equation}
\{\Gamma,H\}=0,
\label{eq:chirala}
\end{equation}
or
\begin{equation}
\{\Gamma,\gamma_i\}=0.
\label{eq:chiralb}
\end{equation}
\end{subequations}%
We note that from two of the three relations (\ref{eq:TRS})-(\ref{eq:chiral})
follows the third relation where the transformation operator
is the product of those from the two relations.
For example, if $H$ has both TRS and PHS,
then $H$ has a chiral symmetry with $\Gamma=TC$.
The antiunitary operators $T$ and $C$ square to plus or minus identity
operator.

\begin{table}[tb]
\begin{center}
\caption{\label{table: real AZ classes}
Real symmetry classes and their
topological classification.
The eight real symmetry classes are characterized
by the presence or the absence of 
time-reversal symmetry ($T$) and particle-hole symmetry ($C$).
Their presence is labeled by the sign of squared operator,
$T^2$ or $C^2$,
and their absence is indicated by 0.
For each class, the parameters $(p,q)$ specifying the relevant real
Clifford algebra, real classifying space $R_{q-p}$ and its zeroth
homotopy group at $d=0$ are listed.
\\
}
\begin{tabular}[t]{ccccccc}
\hline \hline
class & $T$ & $C$ & $\Gamma$ & $(p,q)$ & $R_{q-p}$ & $\pi_0(R_{q-p})|_{d=0}$ \\
\hline
AI   & $+1$~ & 0    &  0     & ~$(d+2,2)$ & ~$R_{0-d}$ & $\mathbb{Z}$   \\
BDI  & $+1$~ & $+1$~ &  1     & ~$(d+1,2)$ & ~$R_{1-d}$ & $\mathbb{Z}_2$ \\
D    & 0    & $+1$~ &  0     & ~$(d,2)$   & ~$R_{2-d}$ & $\mathbb{Z}_2$ \\
DIII & $-1$~ & $+1$~ &  1     & ~$(d,3)$   & ~$R_{3-d}$ & 0              \\
AII  & $-1$~ & 0    &  0     & ~$(d,4)$   & ~$R_{4-d}$ & $\mathbb{Z}$   \\
CII  & $-1$~ & $-1$~ &  1     & ~$(d+3,0)$ & ~$R_{5-d}$ & 0              \\
C    & 0    & $-1$~ &  0     & ~$(d+2,0)$ & ~$R_{6-d}$ & 0              \\
CI   & $+1$~ & $-1$~ &  1     & ~$(d+2,1)$ & ~$R_{7-d}$ & 0              \\
\hline \hline
\end{tabular}
\end{center}
\end{table}

Among the ten AZ symmetry classes,
two symmetry classes (A and AIII) which have neither TRS nor PHS
are called complex classes, and the other eight symmetry classes are
called real classes shown in Table~\ref{table: real AZ classes}.
For a more detailed introduction to the classification of single-particle
Hamiltonians into the ten AZ classes,
we refer the reader to Sec.~1.1 in Ref.~\onlinecite{ryu-njp10}.

It is known that, in each spatial dimension, five out of the
ten AZ symmetry classes have topologically distinct gapped phases,
which are characterized by either an integer ($\mathbb{Z}$) or a binary
($\mathbb{Z}_2$) topological index.\cite{schnyder-ryu08,kitaev09,ryu-njp10}
For example, gapped phases of minimal Dirac models can be topologically
classified by examining how many distinct mass terms 
the Dirac Hamiltonians can have
under given symmetry constraints.\cite{morimoto-clifford13}
When $H$ can accommodate more than one mass terms,
$m_1^{}\gamma_{0,1}^{}+\ldots+m_n^{}\gamma_{0,n}^{}$,
where $\gamma_{0,i}$'s anticommute with each other and with
other $\gamma_i$'s ($i=1,\ldots,d$),
all the gapped ground states of $H$ with different values of
$\bm{m}=(m_1,\ldots,m_n)\ne0$
are adiabatically connected without closing a gap.
This means that the gapped phase of $H$ is topologically trivial.
On the other hand, when $H$ has only a unique mass term ($m\gamma_0$),
the ground state of $H$ with positive $m$ (denoted by $H_+$),
and the ground state of $H$ with negative $m$ (denoted by $H_-$),
are topologically distinct gapped states
separated by a critical point at $m=0$; i.e.,
the two gapped phases with opposite signs of $m$ are
topologically distinct phases.
Distinction between the two classifications, $\mathbb{Z}$ and $\mathbb{Z}_2$,
becomes clear, when we consider a doubled system $H \tensor \sigma_0$,
where $\sigma_0$ is a unit $2\times2$ matrix.
For gapped phases with $\mathbb{Z}_2$ classification, we find an extra
mass term in the doubled system, with which we can find a continuous
deformation 
from $H_+\tensor\sigma_0$ to $H_-\tensor\sigma_0$ or vice versa.
For gapped phases with $\mathbb{Z}$ classification, on the other hand,
we do not find an extra mass term in the doubled system;
in this case topological indices of gapped ground states
of $H_\pm$ can add up.

The classification of mass terms in minimal Dirac models described above
can be systematically performed by considering an extension problem
of Clifford algebra,\cite{kitaev09} as briefly summarized below.
For more details see, e.g., Sec.~III of Ref.~\onlinecite{morimoto-clifford13}.
In this paper we are concerned with real symmetry classes.
For each real AZ class and each spatial dimension $d$,
we can define a real Clifford algebra
$Cl_{p,q}$ (as described later in this section)
whose generators are symmetry operators
(such as $T$, $C$, or $\Gamma$) and kinetic gamma matrices
($\gamma_i$, $i=1,\ldots,d$).
We take a real representation of sufficiently large matrix dimension
for Clifford algebras.
We then examine the possibility of extending a given Clifford algebra
(with a fixed representation) by adding a mass term $\gamma_0$ to the set
of generators of the Clifford algebra.
A set of possible representations of $\gamma_0$
form a manifold called a classifying space.\cite{kitaev09}
The classifying space for the extension
$Cl_{p,q}\to Cl_{p,q+1}$ is given by $R_{q-p}$,
whose explicit form can be found
in literatures.\cite{kitaev09,morimoto-clifford13}
The relation $R_{n+8}=R_n$ is known to hold
(the Bott periodicity).\cite{karoubi,kitaev09}
The topological classification of the Dirac mass terms is then
obtained from the connectivity of the classifying space, i.e.,
the zeroth homotopy group of the classifying space.
The last columns of
Table~\ref{table: real AZ classes} show topological
classification for the eight real symmetry classes at $d=0$.
The classification in $d$ dimensions is obtained by using
the Bott periodicity.\cite{kitaev09}
We note that insulators (superconductors) characterized by
a nontrivial topological index discussed above are called
strong TIs (TSCs),
which should be distinguished from weak TIs (TSCs)
discussed in the next section.

In the rest of this section we give a list of Clifford algebras
and their extension problems which are used to obtain the above-mentioned
classification of strong TIs and TSCs
for the eight real symmetry classes and which will serve as a basis
for the discussion in the following sections.
To this end, we first introduce a real Clifford algebra $Cl_{p,q}$,
which is a $2^{p+q}$-dimensional
real linear algebra generated
by a set of generators $\{e_1,e_2,\ldots,e_{p+q}\}$ satisfying
the algebraic relations
\begin{subequations}
\begin{equation}
\{e_i,e_j\}=0~~\mbox{for}~i\ne j
\end{equation}
and
\begin{equation}
e_i^2=\left\{\begin{array}{ll}
-1, & 1\le i\le p, \\
+1, & p+1\le i\le p+q.
\end{array}\right.
\end{equation}
\end{subequations}
We also introduce an operator $J$ which
plays a role of the imaginary
unit ``$\, i \,$'' in real algebras and
obeys the relations
\begin{equation}
J^2=-1,
\qquad
\{T,J\}=\{C,J\}=[\gamma_i,J]=0.
\end{equation}

\textit{Symmetry classes C and D}.
These two classes have only a PHS:
$C^2=-1$ in class C and $C^2=+1$ in class D.
We define a Clifford algebra $Cl_{p,q}$ generated by
the operators
\begin{subequations}
\begin{align}
\{ C, CJ, J\gamma_1, \ldots, J\gamma_d \},
\end{align}
where $p$ and $q$ are
listed in Table~\ref{table: real AZ classes}.
For topological classification we consider extending $Cl_{p,q}$ to
$Cl_{p,q+1}$ with the generators
\begin{equation}
\{ \gamma_0, C, CJ, J\gamma_1, \ldots, J\gamma_d \}.
\label{eq:algebra for C}
\end{equation}
\end{subequations}

\textit{Symmetry classes BDI, CI, CII, and DIII}.
These four classes have both TRS and PHS.
We have a Clifford algebra $Cl_{p,q}$ generated by
\begin{subequations}
\begin{align}
\{ C, CJ, TCJ, J\gamma_1, \ldots, J\gamma_d \},
\end{align}
which is to be extended to $Cl_{p,q+1}$ generated by
\begin{align}
\{ \gamma_0, C, CJ, TCJ, J\gamma_1, \ldots, J\gamma_d \},
\label{eq:algebra for T and C}
\end{align}
\end{subequations}
where $(p,q)$ are listed for each class in Table~\ref{table: real AZ classes}.

\textit{Symmetry classes AI and AII}.
These classes have a TRS only.
We define a Clifford algebra $Cl_{p',q'}$ generated by
\begin{subequations}
\begin{align}
\{ T, TJ, \gamma_1, \ldots, \gamma_d \},
\end{align}
where $p'$ and $q'$ denote the numbers of generators squaring to
$-1$ and $+1$, respectively;
\begin{equation}
(p',q')=
\left\{\begin{array}{ll}
(0,d+2), & \mbox{AI}, \\
(2,d), & \mbox{AII}.
\end{array}\right.
\end{equation}
We consider extending $Cl_{p',q'}$ to $Cl_{p'+1,q'}$,
with the generators
\begin{align}
\{ J\gamma_0, T, TJ, \gamma_1, \ldots, \gamma_d \}.
\label{eq:algebra for T}
\end{align}
\end{subequations}
Instead of directly studying the extension problem
$Cl_{p',q'}\to Cl_{p'+1,q'}$,
we make use of the isomorphism\cite{kitaev09,morimoto-clifford13}
$Cl_{p',q'}\tensor Cl_{0,2} \simeq Cl_{q',p'+2}$
by tensoring redundant degrees of freedom $Cl_{0,2}\simeq \mathbb{R}(2)$,
to obtain the equivalent extension problem
\begin{align}
Cl_{p,q} \to Cl_{p,q+1}, \qquad (p,q)=(q',p'+2),
\label{eq:modified extension for T only}
\end{align}
with $(p,q)$ listed in Table~\ref{table: real AZ classes}.
Here $\mathbb{R}(2)$ is an algebra of 2 by 2 matrices.

In the following sections we will study the stability of surface states of
weak $\mathbb{Z}_2$ TIs and TSCs,
using minimal Dirac models and Clifford algebras.
We will make use of the one-to-one correspondence
between the existence of
a single mass (multiple masses) in a minimal
Dirac Hamiltonian and nontrivial (trivial) topology of
the corresponding classifying space.

\section{General theory based on Clifford algebras}

\begin{table}[tb]
\begin{center}
\caption{\label{table: weak TI}
Classification of mass terms and existence condition of an
additional kinetic term in 
weak topological insulators and superconductors.
They are determined by the value of $q-p$,
where $p$ and $q$ are the numbers of generators squaring to $-1$
and $+1$ of the real Clifford algebra
which is specified by the symmetry class and the spatial dimension as
listed in Table~\ref{table: real AZ classes}.
The column with $\gamma_0$ shows classification of a mass term $\gamma_0$.
The column with $\gamma_{d+1}$ shows the existence condition of
the kinetic term along the $(d+1)$th direction,
where 0 indicates the existence of such a term, and
both $\mathbb{Z}$ and $\mathbb{Z}_2$ mean the absence.
The last column with $\tilde \gamma_0$ shows classification of
the mass term $\tilde \gamma_0$,
where $\mathbb{Z}$ or $\mathbb{Z}_2$ denotes uniqueness of
the mass term gapping the surface states of a weak topological insulator
or superconductor,
while 0 indicates the existence of multiple mass terms.
\\
}
\begin{tabular}[t]{cccc}
\hline \hline
$q-p$  & $\gamma_0$  & $\gamma_{d+1}$ & $\tilde \gamma_0$  \\ \cline{2-4}
(mod 8) &$\pi_0(R_{q-p})$&$\pi_0(R_{p-q})$& $\pi_0(R_{q-p})$  \\
\hline
0    & $\mathbb{Z}$   & $\mathbb{Z}$ & $\mathbb{Z}$   \\
1    & $\mathbb{Z}_2$ & 0            & $\mathbb{Z}_2$ \\
2    & $\mathbb{Z}_2$ & 0            & $\mathbb{Z}_2$ \\
3    & 0              & 0            & 0              \\
4    & $\mathbb{Z}$   & $\mathbb{Z}$ & $\mathbb{Z}$   \\
5    & 0              & 0            & 0            \\
6    & 0              &$\mathbb{Z}_2$& 0            \\
7    & 0              &$\mathbb{Z}_2$& 0            \\
\hline \hline
\end{tabular}
\end{center}
\end{table}

In this section we study surface states of $(d+1)$-dimensional
weak $\mathbb{Z}_2$ TIs and TSCs which are
stacked layers of $d$-dimensional strong $\mathbb{Z}_2$ TIs and TSCs.
We will show that the $d$-dimensional Dirac Hamiltonian
for the surface states admits only a single mass term,
which corresponds to dimerization of the stacked layers.
The dimerized insulating states with a finite mass
are labeled by a $\mathbb{Z}_2$ index (the sign of the mass).
We will further argue that the surface states are not localized
by disorder as long as disorder average of the dimerization
mass term vanishes.
We show these in several steps below.

Let us consider a $d$-dimensional strong $\mathbb{Z}_2$ TI or TSC
described by the Hamiltonian $H$ in Eq.~(\ref{eq:Dirac-H}).
The gapped ground state of $H$ has a nontrivial $\mathbb{Z}_2$
topological index in accordance with $\pi_0(R_{q-p})=\mathbb{Z}_2$
for $q-p=1,2$ (mod 8), where $R_{q-p}$ is the classifying space
associated with the extension problem
\begin{align}
Cl_{p,q} \to Cl_{p,q+1}
\end{align}
with $(p,q)$ listed in Table~\ref{table: real AZ classes}
for each symmetry class; see also Table~\ref{table: weak TI}.
In the following discussions where we explain our theory
based on Clifford algebras,
we will use, as examples, the Clifford algebras defined
in Eq.~(\ref{eq:algebra for T and C}) for
time-reversal symmetric TSCs in class BDI, CI, CII, or DIII.
The same theory can be directly applied to classes C and D
which have a PHS only,
since their relevant Clifford algebras (\ref{eq:algebra for C}) are
obtained by just dropping $TCJ$ from Eq.~(\ref{eq:algebra for T and C}).
It is also applicable to classes AI and AII with TRS only,
since their equivalent extension problems
(\ref{eq:modified extension for T only})
have the same mathematical structure.

We describe gapless surface states of $d$-dimensional strong
$\mathbb{Z}_2$ TIs and TSCs as domain wall states of
the massive Dirac Hamiltonian (\ref{eq:Dirac-H}).
Namely, we assume that the Dirac mass $m$ is a function of $x_d$
and changes its sign at $x_d=0$ (a kink).
This yields $(d-1)$-dimensional gapless surface Dirac fermions
localized at $x_d=0$.
The wavefunction of the surface Dirac fermions can be written as
$\Psi(x_1,\ldots,x_{d-1})\psi(x_d)$, where $\Psi(x_1,\ldots,x_{d-1})$
is an eigenfunction of the $(d-1)$-dimensional surface Dirac Hamiltonian
$H_{d-1}=-i\sum^{d-1}_{j=1}\gamma_j\partial_{x_j}$, and the
localized wavefunction $\psi(x_d)$ is a solution to the equation
$[-i\gamma_d \partial_{x_d} + m(x_d) \gamma_0]\psi(x_d)=0$, i.e.,
\begin{align}
\psi(x_d)=\exp \!\left[-i \int^{x_d}_0 dx_d'  m(x_d') \gamma_d \gamma_0 \right]
|n \rangle,
\end{align}
where $|n \rangle$ is chosen from eigenvectors of $-i\gamma_d\gamma_0$
(eigenvalue $+1$ or $-1$)
such that $\psi(x_d)$ is normalizable.
We note that $-i\gamma_d\gamma_0$ commutes with the gamma matrices
$\gamma_i$ ($i=1,\ldots,d-1$) and symmetry operators in the Clifford
algebra $Cl_{p,q+1}$.
This is formally written as 
$Cl_{p,q+1} \simeq Cl_{p-1,q} \tensor Cl_{1,1} 
\simeq Cl_{p-1,q} \tensor \mathbb{R}(2)$.
The $\mathbb{R}(2)$ degrees of freedom, which correspond to the localized
wavefunction $\psi$, should be kept intact in the following procedures
of building weak TIs (TSCs)
by stacking ``layers'' of $d$-dimensional TIs (TSCs)
in the $(d+1)$th direction.
We set the ``interlayer'' spacing to unity for simplicity.

As we show later, there always exists a kinetic gamma matrix $\gamma_{d+1}$
that satisfies the symmetry constraints
[Eqs.~(\ref{eq:TRS}) and (\ref{eq:PHS})] and anticommutes with
all the other gamma matrices, $\{\gamma_{d+1}, \gamma_i\}=0$ ($i=0,\ldots,d$).
We can use the symmetry allowed operator $\gamma_{d+1}$
for ``inter-layer'' coupling between the
$(d-1)$-dimensional Dirac surface states on neighboring ``layers''.
The $d$-dimensional surface states of the $(d+1)$-dimensional
weak TI (TSC) are then governed by
the Schr\"odinger equation
\begin{align}
-i\sum_{j=1}^{d-1} \gamma_j \frac{\partial\Psi_l}{\partial x_j}
-\frac{i}{2} t \gamma_{d+1}(\Psi_{l+1}-\Psi_{l-1})
=
E\Psi_l,
\label{eq:stacking}
\end{align}
where $\Psi_l(x_1,\ldots,x_{d-1})$ is the wavefunction of
surface Dirac fermions in the $l$th layer,
and $t$ is an interlayer hopping matrix element.
Note that the interlayer hopping term is compatible with TRS and PHS.
In the momentum space the $d$-dimensional surface Hamiltonian reads
\begin{align}
H_{d+1}^{(0)}= \sum_{i=1}^{d-1} \gamma_i k_i  + t \gamma_{d+1} \sin k_{d+1}.
\end{align}
After linearizing the dispersion near $k_{d+1}=0$ and $\pi$,
we have two surface Dirac cones centered at
$(k_1,\ldots,k_{d-1},k_{d+1})=(0,\ldots,0,0)$ and $(0,\ldots,0,\pi)$.
While we have chosen a particular stacking structure
in Eq.~(\ref{eq:stacking}),
this should suffice to discuss properties of weak TIs/TSCs
that generally possess two surface Dirac cones.

We now prove the existence of $\gamma_{d+1}$.
Let us look at the extension problem of
$Cl_{p-1,q+1}\to Cl_{p,q+1}$,
which is written in terms of generators as
\begin{align}
&\{\gamma_0,C,CJ,TCJ,J\gamma_1,\ldots,J\gamma_{d-1}\}\n
&~
\to
\{\gamma_0,C,CJ,TCJ,J\gamma_1,\ldots,J\gamma_{d-1},J\gamma_d\},
\label{eq:existence gamma d+1}
\end{align}
whose classifying space is $R_{p-q}$.
The topological classification for this extension of adding $\gamma_d$
is trivial, $\pi_0(R_{p-q})=0$, for $q-p=1,2$
as shown in Table~\ref{table: weak TI}.
This implies that,
when we try to add one kinetic gamma matrix $\gamma_d$ to a set of $\gamma_i$'s
$(i=0,\ldots,d-1)$, we can always find another kinetic gamma matrix, $\gamma_{d+1}$,
that anticommutes with the other $\gamma_i$'s and is compatible with
symmetry constraints.
In passing, we note that, when the $d$-dimensional TI or TSC
is characterized by an integer topological
index $\mathbb{Z}$ ($Cl_{p,q}\to Cl_{p,q+1}$ with $q-p=0,4$),
the zeroth homotopy group of the classifying space $R_{p-q}$
for the extension problem (\ref{eq:existence gamma d+1}) is 
also $\mathbb{Z}$ (see Table \ref{table: weak TI}),
implying that we cannot find $\gamma_{d+1}$ to have
surface Dirac cones;
instead, we have a chiral metallic surface
as in the case of the two-dimensional
surface of layers of 2-dimensional integer quantum Hall states.\cite{chalker}
The surface states can acquire finite dispersion along the $k_{d+1}$
direction from interlayer hopping operators which do not
anticommute with the $(d-1)$-dimensional Dirac Hamiltonian $H_{d-1}$.

We introduce a grading of 2 by 2 matrix $\tau_j$ to distinguish
the two Dirac cones
($\tau_z=+1$ for $k_{d+1}=0$ and $\tau_z=-1$ for $k_{d+1}=\pi$)
on the $d$-dimensional surface
and rewrite the surface Dirac Hamiltonian as
\begin{align}
H_{d+1}=\sum_{i=1}^{d-1} \gamma_i \tau_0 k_i  + t \gamma_{d+1} \tau_z k_{d+1}.
\end{align}
It is easy to see that there is only a single mass term $M\gamma_{d+1}\tau_y$
which can be added to $H_{d+1}$,
as we show using Clifford algebras below.
In fact, another candidate mass term $\gamma_{d+1}\tau_x$ is not
allowed by TRS and PHS [Eqs.~(\ref{eq:TRSb}) and (\ref{eq:PHSb})]
(here we have assumed
$[\mathcal{K},\tau_x]=\{\mathcal{K},\tau_y\}=[\mathcal{K},\tau_z]=0$
and $\{T,\gamma_{d+1}\}=[C,\gamma_{d+1}]=0$).
The mass term $\gamma_{d+1}\tau_y$ gaps out the two surface
Dirac cones and is compatible with TRS and PHS.
Physically, this mass term corresponds to dimerization of the ``interlayer''
hopping, as one can see from the fact that the translation in the
$x_{d+1}$ direction ($l\to l+1$) corresponds to an operation of $\tau_z$,
and this mass term breaks the translation symmetry
as $\{\tau_z,\gamma_{d+1}\tau_y\}=0$.\cite{FuKane2012}

The uniqueness of the dimerization mass term
$\tilde \gamma_0=\gamma_{d+1}\tau_y$
is understood by considering the extension problem
$Cl_{p+1,q+1} \to Cl_{p+1,q+2}$, i.e.,
\begin{align}
&\{\gamma_0,C,CJ,TCJ,J\gamma_1,\ldots,J\gamma_d,J\gamma_{d+1}\tau_z\} \n
&~\to
\{\tilde \gamma_0, \gamma_0,C,CJ,TCJ,J\gamma_1,\ldots,
 J\gamma_d,J\gamma_{d+1}\tau_z\},
\end{align}
whose classifying space is again $R_{q-p}$ with $\mathbb{Z}_2$
classification, $\pi_0(R_{q-p})=\mathbb{Z}_2$
(Table \ref{table: weak TI}).
Here we have included the original mass term $\gamma_0$ and all
the kinetic gamma matrices $\gamma_i$ ($i=1,\ldots,d+1$) in the
Clifford algebra $Cl_{p+1,q+1}$ to be extended, because we are seeking
an extra mass term under the fixed representation of those gamma matrices
and symmetry operators.
Hence the dimerization term that we have found,
$\tilde \gamma_0=\gamma_{d+1}\tau_y$,
is the unique mass term to gap out the surface Dirac cones of weak
$\mathbb{Z}_2$ TIs and TSCs.
The gapped $d$-dimensional surface is a (strong) $\mathbb{Z}_2$
TI or TSC.

Finally, let us discuss Anderson localization of the $d$-dimensional surface
states of $(d+1)$-dimensional weak $\mathbb{Z}_2$ TIs and TSCs.
We assume that disorder is weaker than the bulk band gap
and changes slowly in space.
The disorder potential gives rise to random signs of the dimerization
mass $M$.
The surface is then split into gapped regions (domains) of different
$\mathbb{Z}_2$ indices, and there appear gapless helical states
propagating along the domain boundaries.
When we assume uniformity of the disorder-averaged surface and
a vanishing mean of the dimerization mass term $M\tilde \gamma_0$,
we expect that helical domain-wall states should percolate
throughout the surface and never be localized.
This mechanism is indeed at work for the metallic phase separating
two insulating phases with distinct $\mathbb{Z}_2$ topological
indices
in the phase diagram of disordered two-dimensional insulators
in class AII.\cite{Obuse2007}
A similar physics is known in the integer quantum Hall effect of Dirac
fermions, where an unstable critical point between quantum Hall
plateaus $\sigma_{xy}=\pm e^2/2h$ is realized under random magnetic
fields and random mass, both with zero mean.\cite{Nomura2008}
We thus conclude that, even in the presence of disorder, the surface states
of weak $\mathbb{Z}_2$ TIs and TSCs are not
localized and remain either metallic or critical,
as long as the disorder average of the dimerization mass term vanishes.
This conclusion is a natural generalization of
the stability of surface states of three-dimensional weak $\mathbb{Z}_2$
topological insulators which was a subject of active research
recently.\cite{Ringel,Mong,FuKane2012,Obuse}

\section{Examples}

In this section we consider three examples of weak $\mathbb{Z}_2$
topological insulators.

We start with a three-dimensional weak TI of class AII,
which is a stack of two-dimensional strong $\mathbb{Z}_2$
TIs.\cite{Mong,FuKane2012}
Each TI layer is described as
\begin{align}
H_\t{2D}= k_x \sigma_x \tau_x + k_y \tau_y + m(y) \tau_z,
\label{H_2D}
\end{align}
where $\sigma_i$ and $\tau_i$ are Pauli matrices ($i=x,y,z$).
The Hamiltonian has a TRS with $T=i\sigma_y \mathcal{K}$.
The mass $m(y)$ is assumed to have a kink, where
a helical edge mode is formed as one of the eigenstates of
$\tau_x=-i \tau_y \tau_z$.
Here we take $\tau_x=+1$ without loss of generality.

We stack two-dimensional TI layers along the $z$ direction
to build a three-dimensional weak TI.
The interlayer hopping term for the helical edge states
is given by $t\sigma_y\tau_x$,
which anticommutes with the gamma
matrices in Eq.\ (\ref{H_2D}) and $T$.
The effective Hamiltonian for the surface Dirac fermions
of the three-dimensional weak TI then reads
\begin{align}
H_\t{3D}= k_x \sigma_x + t\sin k_z \sigma_y,
\end{align}
where we have suppressed $\tau_x=+1$,
since the $\tau$ sector is fixed in the helical edge states
forming the surface Dirac cones.
The original $\mathbb{Z}_2$ classification of two-dimensional TIs
implies that there is no extra mass term in this two-dimensional
representation with $\sigma_i$.
However, since we have two valleys of Dirac cones at $k_z=0$ and $\pi$,
which we denote by $\rho_z=+1$ and $-1$ with another set of
Pauli matrices $\rho_i$ ($i=x,y,z$),
we can find the dimerization mass term $\tilde{m}\sigma_y\rho_y$
which gaps out the surface Dirac cones.
This yields
the massive surface Dirac Hamiltonian
\begin{align}
H_\t{3D}= k_x \sigma_x + tk_z \sigma_y \rho_z + \tilde m \sigma_y \rho_y.
\label{H_3D AII}
\end{align}
One can easily verify that $\tilde{m}\sigma_y\tau_x\rho_y$ is
the unique mass term
that is invariant under $T=i\sigma_y\mathcal{K}$ and anticommutes
with all the gamma matrices
($\tau_z$, $\sigma_x\tau_x$, $\tau_y$, $\sigma_y\tau_x\rho_z$).
The uniqueness of the mass term means that
the gapless surface states are neither gapped nor localized
unless the translation symmetry in the layer direction is broken
by dimerization.
We note that the uniqueness of the mass term in the Dirac Hamiltonian
in Eq.\ (\ref{H_3D AII}) was already pointed out in
Refs.~\onlinecite{Mong} and \onlinecite{FuKane2012}.
Mong \textit{et al.}\ studied numerically the Anderson localization
of this Hamiltonian with additional time-reversal invariant disorder
and showed that the critical point $\tilde{m}=0$ turns into a metallic
phase separating two dimer insulating phases.\cite{Mong}

Next we discuss two examples of two-dimensional weak $\mathbb{Z}_2$ TSCs
which are formed as a multi-leg ladder of superconducting wires.
The first of these is a weak two-dimensional TSC in class D.
The BdG Hamiltonian for a one-dimensional spinless $p$-wave superconductor
can be written as
\begin{align}
H_\t{1D}=k_x \sigma_x + m(x) \sigma_z,
\label{eq:weak-H1d-D}
\end{align}
where $\sigma$ denotes particle-hole grading.
The Hamiltonian has a PHS ($C=\sigma_x \mathcal{K}$).
We introduce a kink in the mass $m(x)$ (which is nothing but
the chemical potential of the electrons in the wire) to
obtain a Majorana bound state localized at the kink,
which is an eigenstate of $\sigma_y=-i\sigma_z\sigma_x$.
We can set $\sigma_y=+1$ for the bound state for simplicity.
We find that the same operator $\sigma_y$ can be used as
an interchain hopping operator which commutes with $C$ and
anticommutes with $\sigma_x$ and $\sigma_z$.
A Majorana bound state from each chain is coupled by
the interchain hopping term and acquires
a kinetic term $t\sigma_y \sin k_y$, where $k_y$ is the momentum
along the direction perpendicular to the chain direction $x$.
The Majorana edge states thus formed have two Dirac points
($k_y=0$ and $\pi$), which we denote by $\tau_z=+1$ and $-1$.
We find that the Majorana edge states of the two-dimensional
weak TSC are governed by the Hamiltonian
\begin{align}
H_\t{2D} = t k_y \sigma_y \tau_z + \tilde m \sigma_y \tau_y,
\end{align}
where $\tilde{m}\sigma_y\tau_y$ is the unique mass term
which anticommutes with $C$ and all the gamma matrices
($\sigma_z$, $\sigma_x$, $\sigma_y\tau_z$).
In the presence of disorder the Majorana edge states remain critical
(i.e., diverging localization length) as long as the disorder
average of the dimerization mass term vanishes.\cite{BFGM}
Incidentally, we note that, if we impose a chiral symmetry on
this model (which is now in class BDI),
the Majorana bound states form a stable flat band at zero energy
as follows.
In a TSC of class BDI which is described by the Hamiltonian
(\ref{eq:weak-H1d-D}) with a chiral symmetry $\Gamma=\sigma_y$,
Majorana bound states are again an eigenstate of $\sigma_y$.
However, the interchain hopping operator $\sigma_y$ is not available anymore,
if we impose the chiral symmetry $\Gamma=\sigma_y$.
Hence the Majorana end states remain at zero energy as chiral zeromodes
and cannot be gapped,
which is in agreement with the fact that one-dimensional BDI TSCs
are characterized by an integer ($\mathbb{Z}$) topological index.

The last example we discuss in this section
is a two-dimensional weak $\mathbb{Z}_2$ TSC in class DIII.
We begin with a one-dimensional strong TSC with the Hamiltonian
\begin{align}
H_\t{1D}=k_x \sigma_z \tau_z + m(x) \tau_x,
\label{H_1D DIII}
\end{align}
where we have TRS $T=i\sigma_y \mathcal{K}$
and PHS $C=\sigma_y\tau_y \mathcal{K}$.
This Hamiltonian was discussed in Ref.~\onlinecite{zhang-kane-tmsc}
as a model of
Rashba wires in proximity to a $s_\pm$-wave superconductor,
where we can regard the first term as a spin-orbit coupling and
the second as a superconducting pair potential
with $\sigma$ and $\tau$ spanning spin and particle-hole degrees of freedom,
respectively.
At each end of a one-dimensional TSC, a Kramers pair of Majorana bound
states appear, and they are an eigenstate of $\sigma_z\tau_y$.
We may take the sector of $\sigma_z\tau_y=+1$ in the following discussion.
When we stack one-dimensional chains of TSCs along the $y$ direction,
we can take $\sigma_x \tau_z \sin k_y$ as a kinetic term due to
interchain hopping.
Denoting the two Dirac points at $k_y=0$ and $\pi$ by $\rho_z=+1$ and $-1$,
we have the effective Hamiltonian for the Majorana edge states
of the two-dimensional weak TSC,
\begin{align}
H_\t{2D}=k_y \sigma_x \tau_z \rho_z + \tilde m \sigma_x \tau_z \rho_y,
\end{align}
where $\tilde m \sigma_x\tau_z\rho_y$ is again the unique mass term
which is compatible with TRS and PHS,
and anticommutes with gamma matrices $\tau_x$, $\sigma_z\tau_z$,
and $\sigma_x\tau_z\rho_z$.
The Majorana edge states should remain critical as long as the
edge is uniform on average, i.e., when the dimerization term is absent
after disorder average.\cite{BFGM}

\section{Summary}

We have discussed the surface stability of weak $\mathbb{Z}_2$
TIs (TSCs) which are stacked layers of strong $\mathbb{Z}_2$ TIs (TSCs),
by examining the topological structure of the surface Dirac Hamiltonians,
using Clifford algebras.
We have shown the uniqueness of a Dirac mass term which causes scattering
between the two surface Dirac cones and gaps them out by inducing dimerization
of stacked layers.
The dimerized insulating surface is a strong TI (TSC)
with a $\mathbb{Z}_2$ index determined by the sign of the unique Dirac mass.
Thus the point where the Dirac mass vanishes is a quantum critical point,
which either remains to be critical or becomes metallic in the presence
of disorder potential with a vanishing mean, i.e., without dimerization.

We note that our discussion based on Dirac Hamiltonians
is valid only for small energy scale,
where we can linearize the dispersion of gapless surface states.
At larger energy scale one should include a quadratic momentum
dependence and take into account renormalization of a Dirac mass.\cite{groth}
Furthermore, when disorder strength is large and comparable
with a bulk energy gap,
the description with Dirac Hamiltonians is no longer appropriate.
In that case, we have to consider original lattice Hamiltonians.
Indeed numerical calculations on lattice models have shown that
an insulating phase appears at strong disorder.\cite{Kobayashi13,Yamakage13}

Finally, we point out that the physics we discussed here is also
relevant for topological crystalline insulators
which have gapless surface states
protected by spatial symmetries (such as a mirror symmetry).\cite{hsieh12,chiu13,ueno2013symmetry,morimoto-clifford13}
An interesting question we may ask is the stability of
the gapless surface states when the required spatial symmetry
is retained only on average.\cite{Yao-Ryu13,fulga2012statistical}
Using Clifford algebras, we can study stability of these gapless
surface states against disorder which breaks spatial symmetries,
which will be reported elsewhere.\cite{morimoto-in-prep}
This analysis can be applied to
topological crystalline insulators with an average mirror symmetry,
e.g., Pb$_{1-x}$Sn$_x$Te materials where four Dirac cones appear
on the (001) surface.\cite{hsieh12}

\section{acknowledgment}
This work was supported by Grants-in-Aid from the Japan Society for
Promotion of Science (Grants No.~24840047 and No.~24540338)
and by the RIKEN iTHES Project.


\end{document}